# Silicon-integrated scandium-doped aluminum nitride electro-optic modulator

TIANQI XU,[1] YUSHUAI LIU,[3,4,5] YUANMAO PU,[1] YONGXIANG YANG,[1] QIZE ZHONG,[1,2] XINGYAN ZHAO,[1] YANG QIU,[1] YUAN DONG,[1] TAO WU,[3,4,5] SHAONAN ZHENG,[1 *] AND TING HU[1, 2 *]

[1]School of Microelectronics, Shanghai University, Shanghai, 201800, China

[2] Shanghai Key Laboratory of Chips and Systems for Intelligent Connected Vehicle, Shanghai, 200444, China.

[3]School of Information Science and Technology, ShanghaiTech University, Shanghai, 201210, China

[4]Shanghai Institute of Microsystem and Information Technology, Chinese Academy of Sciences, Shanghai, 200050, China

[5]University of Chinese Academy of Sciences, Beijing, 100049, China

E-mail: *hu-t@shu.edu.cn, *snzheng@shu.edu.cn

**Abstract:** Scandium-doped aluminum nitride (AlScN) with an asymmetric hexagonal wurtzite structure exhibits enhanced second-order nonlinear and piezoelectric properties compared to aluminum nitride (AlN), while maintaining a relatively large bandgap. It provides a promising platform for photonic integration and facilitates the seamless integration of passive and active functional devices. Here, we present the design, fabrication, and characterization of AlScN EO micro-ring modulators, introducing active functionalities to the chip-scale AlScN platform. These waveguide-integrated EO modulators employ sputtered AlScN thin films as the light-guiding medium, and the entire fabrication process is compatible with complementary metal oxide semiconductor (CMOS) technology. We characterize the high-frequency performance of an AlScN modulator for the first time, extracting a maximum in-device effective EO coefficient of 2.86 pm/V at 12 GHz. The devices show a minimum half-wave voltage-length product of 3.12 V·cm and a 3-dB modulation bandwidth of approximately 22 GHz. Our work provides a promising modulation scheme for cost-effective silicon-integrated photonics systems.

**Introduction**

Over the past decades, photonic integrated circuits (PICs) have attracted widespread research interest due to their advantages of scalability, low power consumption, and potential for cost-effective mass production[1,2]. PICs hold promise to achieve chip-scale systems with integrated functionalities. Electro-optic (EO) modulators that manipulate the amplitude or phase of light carriers with electrical signals are one of the key components in PICs[3], finding broad applications in high-speed optical communications[4], optical interconnections[5], quantum information processing[6], computing[7] and so on. The escalating demand in these fields has propelled the development of high-performance EO modulators.

Silicon-integrated EO modulators have been widely reported and can be categorized as follows: 1) Silicon (Si) modulators exploiting plasma dispersion effect[8,9]. They often require complex doping architectures and additional thermo-optic tuning heaters while confronting strict modulation speed, efficiency, and carrier absorption trade-offs. 2) Two-dimensional (2D) material-based modulators (e.g., graphene) rely on the Pauli-blocking effect[10]. Due to the complexity of material transfer techniques, wafer-level fabrication is currently not realized and results in substantial cost increases. 3) Ferroelectric oxides (e.g., Lithium Niobate (LN) and Barium titanate (BTO)) modulators utilizing Pockels effect[11–15]. Compared with the plasma dispersion effect, the Pockels effect offers ultrafast, linear, and pure real refractive index modulation across a broad wavelength range without additional absorption losses. Therefore, LN and BTO modulators have been extensively studied in recent years. Unfortunately, lithium niobate-on-insulator (LNOI) thin films on 8-inch silicon substrates realized by the smart-cut method are costly. The preparation conditions for single-crystal BTO thin films on insulator substrates are even more stringent. The prevailing molecular beam epitaxy (MBE) technology suffers from low deposition rates. Furthermore, BTO's Curie temperature is only 125°, which would bring the potential process challenges and make it unsuitable for high-temperature operation. Overall, finding a cost-effective electro-optic material suitable for large-scale integration remains a challenge.

Aluminum nitride (AlN) thin films on the insulator substrate fabricated by a mature complementary metal oxide semiconductor (CMOS)-compatible sputtering process also exhibit Pockels effect. This renders them a compelling EO material platform, offering cost-effective alternatives to LN and BTO. Waveguide-based AlN modulators operating in the communication band have been experimentally demonstrated showing an EO coefficient of 1 pm/V and a modulation speed of 4.5 Gb/s[16]. A recent study has experimentally verified that the optical second-order susceptibility ($\chi^{(2)}$) components $d_{13}$ and $d_{33}$ of scandium-doped aluminum nitride (AlScN) thin films increase significantly with increasing Sc concentration[17]. Specifically, the $d_{33}$ of $Al_{0.9}Sc_{0.1}N$ is enhanced by a factor of 3, while that of $Al_{0.64}Sc_{0.36}N$ is increased by a factor of 12 compared to undoped AlN films. Since the EO coefficient $r_{ijk}$ is proportional to the $\chi^{(2)}$ [18], the Pockels effect is predicted to be enhanced by about 2.7 and 10 times for scandium-doped aluminum nitride films with concentrations of 10% and 36%, respectively. AlScN maintains a relatively wide bandgap[19]. It is also compatible with various substrates such as silicon[20,21], silicon dioxide[22], and sapphire[19], rendering it a promising platform for integrated photonic applications. Supplementary Table 1 lists the detailed comparison of some common electro-optic materials with AlScN, specifically parameters related to high-frequency response, power handling capability, and stability of the modulator.

In this work, we present EO modulators on the AlScN-on-insulator (AlScNOI) platform. Micro-ring resonators (MRRs) are demonstrated to achieve effective EO modulation while minimizing footprint and reducing power consumption. The AlScN films serving as the light-conducting medium are fabricated using a CMOS-compatible sputtering process, providing the feasibility of large-scale manufacturing of EO modulators. The in-device effective EO coefficient of the AlScN is measured to be 1.1 pm/V under direct current (DC) conditions and 2.86 pm/V under high-frequency conditions, and the measured optical bandwidth is up to 22 GHz. Our findings show the potential of AlScN-based modulators for realizing nonlinear optical applications.

**Result and Discussion**

**Passive Devices.** A critical initial step in fabricating an integrated AlScN modulator is to prepare an AlScN thin film with high-quality and well-aligned crystal orientation, since the Pockels effect strongly relies on the relative alignment between the applied electric field, the optical polarization and the crystal orientation. Using radio frequency magnetron reaction sputtering technology, we deposit 400-nm-thick AlScN thin films (with an ordinary refractive index $n_o = 2.13$ and an extraordinary refractive index $n_e = 2.2$ at 1550 nm) on thermally grown silicon dioxide substrate. The atomic percentages of Al and Sc in the film are 90.4% and 9.6%, respectively. In our recently published work[23–25], a symmetric ($2\theta = \theta$) scan of X-ray diffraction (XRD) investigation for the same AlScN film confirms the deposition of the AlScN film along the (0002) crystal orientation, with its c-axis normal to the substrate surface. Additionally, the sharp rocking curve further ensure the consistency in the c-axis orientation of the AlScN film. This c-axis oriented film ensures the maximum utilization of the EO coefficient $r_{33}$ and $r_{13}$.

To maximize the electro-optic interaction, we employ a fully-etched single-mode waveguide design. As illustrated in Figure 1a, a single transverse-electric (TE$_{00}$) optical mode is well confined in the AlScN waveguide with a thickness of 400 nm and a typical width of 1 μm. The calculated effective refractive index $n_{eff}$ at 1550 nm is 1.73. To define the waveguide patterns constituting the MRR, the electron beam lithography (EBL) and Cl$_2$/Ar-based inductively coupled plasma (ICP) reactive ion etching (RIE) process are used (see Methods). Figure 1b presents a micrograph of the fabricated device, a pair of grating couplers is used to couple light into and out of the chip. The scanning electron microscope (SEM) image in Figure 1c displays a sidewall angle of 76° for the etched waveguide cross-section.

The propagation loss of the waveguide is evaluated by the cutback method. Spiral waveguides of different lengths (0.5 cm, 1.2 cm and 1.6 cm) are fabricated. Figure 1d presents the transmission spectra of each spiral waveguide in the wavelength range from 1520 nm to 1580 nm. Through multiple measurements and the error bar method, the averaged fitted curve at 1550 nm is presented in Figure 1e, revealing a propagation loss of 7.45 ± 0.21 dB/cm. The loss is mainly attributed to rough sidewalls and Sc-based non-volatile compounds produced during the etching process[26], which can be significantly reduced through optimizing the etching recipe. Furthermore, the intercept of the fitted curve in Figure 1e is -19.01 ± 0.26 dB, attributed to the coupling losses from a pair of grating couplers and the bending losses of the spiral waveguides.

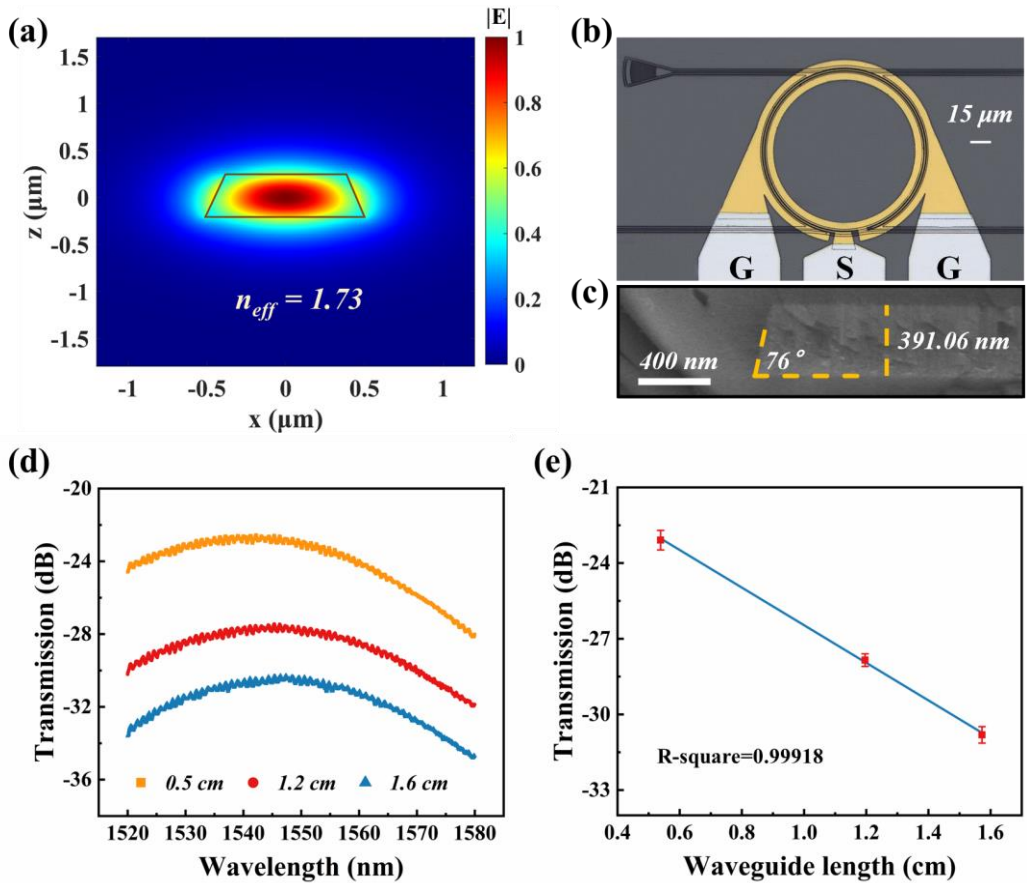

**Fig. 1 a** Simulation of waveguide cross-section with the fundamental TE$_{00}$ mode. **b** Micrograph of the fabricated AlScN micro-ring modulator. **c** SEM image of the AlScN waveguide cross-section. **d** Measured transmission spectra of spiral waveguides of different lengths. The optical transmission is normalized to 0-dBm launched optical power and additional losses in the measurement optical link. **e** Waveguide total insertion loss versus waveguide lengths. The slope of the linear fit indicates the waveguide propagation loss in dB/cm at 1550 nm.

To realize high-performance micro-ring EO modulators, it is necessary to design their structures for critical coupling and determine the suitable quality factor (Q). An excessively high Q will restrict the EO bandwidth due to extended photon lifetime. Consequently, for applications of high-speed modulators, MRRs with a Q of 5000-

25000 are typically used[27]. In addition, a low extinction ratio (ER) reduces the modulation depth and thus degrades the modulation efficiency[28]. We design both add-drop and all-pass resonators utilizing a design approach that combines the transfer-matrix method with the Finite Difference Time Domain (FDTD) numerical simulations. The appropriate coupling coefficients between the straight waveguides and the micro-rings are simulated to satisfy the critical coupling condition. Considering the process tolerance, several sets of MRRs with various gap values around the optimal setting are also fabricated. To characterize the transmission properties of the MRRs, light from the tunable laser (Keysight 8164B) is coupled into/out of grating couplers via polarization-maintaining (PM) fibers. Figure 2a displays the transmission spectrum of the add-drop ring with a diameter of 80 μm. Coupling gaps are 250 nm for the pass port and 300 nm for the drop port, respectively. The free spectral range (FSR) is ~4.4 nm; the maximum ER is 30 dB. The loaded Q value is determined to be 7781 through Lorentz fitting, as depicted in Figure 2b Similarly, Figure 1c illustrates the transmission spectrum of an all-pass ring with a diameter of 200 μm, a gap of 150 nm, and an FSR of ~1.7 nm. It exhibits a maximum ER of 25 dB, with a loaded Q of 20305, as depicted in Figure 1d.

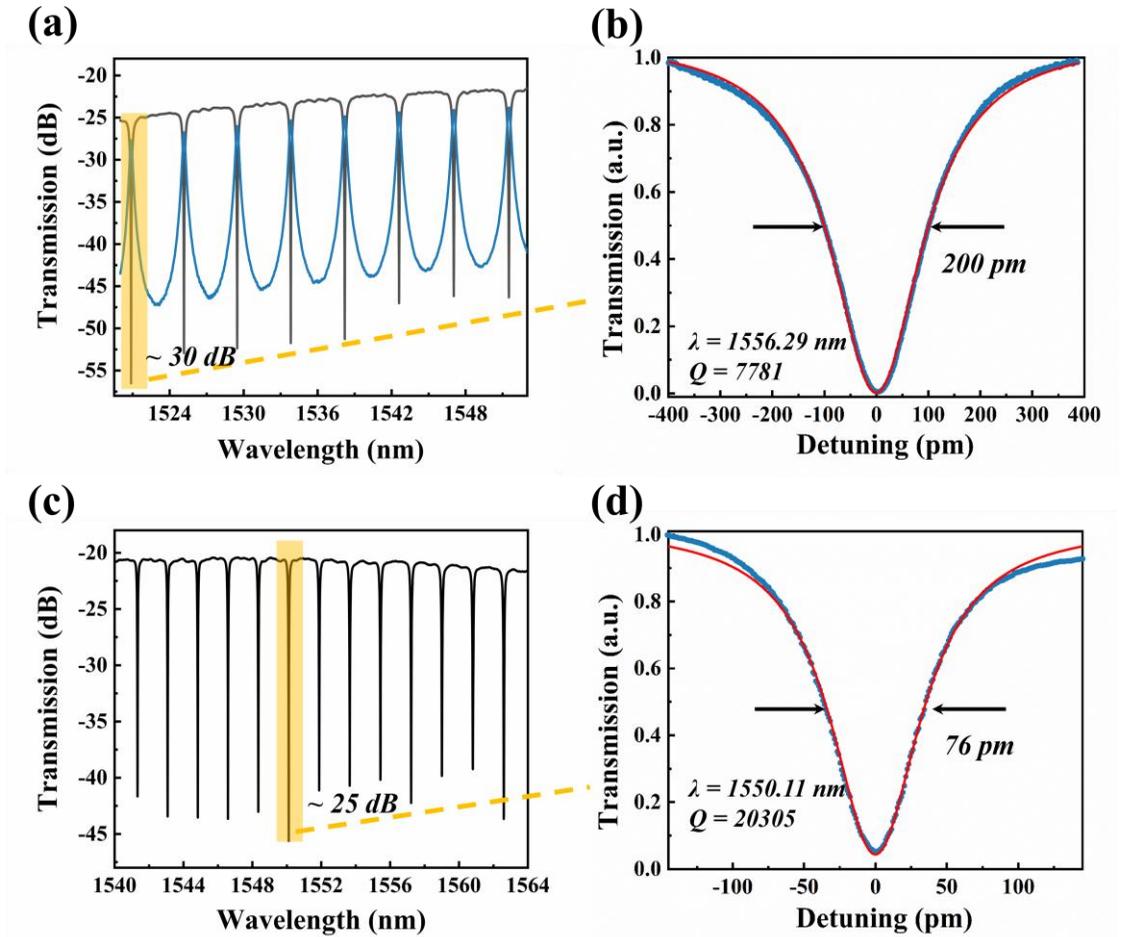

**Fig. 2** Measured transmission spectrum of an **a** add-drop, and an **c** all-pass MRRs. **b d** Normalized resonance profiles fitted according to the Lorentzian function to extract Q values.

**Principle, Design and DC characterization.** When the doping concentration of Sc is 9.6%, the crystalline phase of AlScN exhibits a hexagonal wurtzite structure similar to AlN, both possessing non-centrosymmetric crystal structures. The electro-optic tensor of AlScN has only five non-zero components: $r_{13} = r_{23}$, $r_{33}$, $r_{42} = r_{51}$. Using the refractive index ellipsoid method to describe the Pockels effect [29], the applied electric field must be parallel to the c-axis of the AlScN film to extract the maximum EO coefficients $r_{13}$ and $r_{33}$ (see Methods) [29,30].

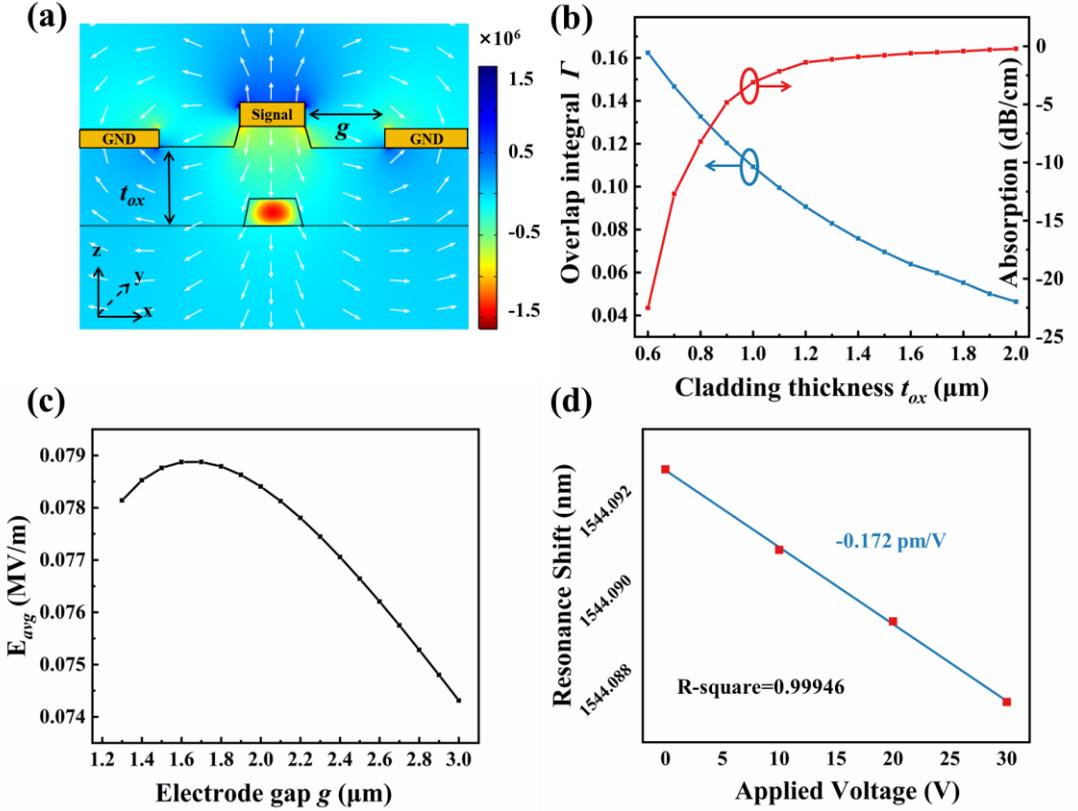

**Fig. 3 a** Cross-sectional view of the numerically calculated electric field distribution. The dominating effective electric field contributing to modulation is the out-of-plane component $E_z$. **b** The numerically calculated electro-optic overlap integral and optical absorption as functions of deposited oxide cladding thickness ($t_{ox}$). With a fixed electrode gap ($g$) of 1.6 μm, the overlap integral ($\Gamma$) decreases as the cladding thickness increases from 0.6 μm to 2 μm. Nevertheless, the optical attenuation caused by the electrodes can be mitigated. **c** The average longitudinal electric field ($E_z$) versus the electrode gap when a voltage of 1 V is applied. In the experiment, the thickness of the $SiO_2$ cladding is chosen to be 1.8 μm to minimize optical absorption while maintaining a suitable electro-optic overlap. The electrode gap is set to 1.6 μm to maximize $E_z$ within the waveguide region. **d** Experimental DC characterization of the EO micro-ring modulator. The resonance wavelength is shifted by ~5.2 pm over a DC voltage range from 0 V to 30 V.

Figure 3a presents the simulated electric field distribution obtained through finite element method (FEM). The electrodes are positioned directly above the waveguide, with ground electrodes located on both sides of the waveguide to maximize the utilization of the out-of-plane electric field ($E_z$). The oxide cladding thickness ($t_{ox}$) (Figure 3b) is firstly optimized by considering the trade-off between the overlap integral ($\Gamma$) that impact modulation efficiency (blue dotted line) and the additional losses associated with the metal electrodes (red dotted line). Subsequently, the signal width above the waveguide is set to 0.2 μm wider than the bottom width of the waveguide to mitigate the influence of signal electrode misalignment during the lithography process. Finally, with the thickness of the oxide cladding $t_{ox}$ is fixed to 1.8 μm, the electrode spacing $g$ between the signal and ground electrodes (Fig. 3c) is set to 1.6 μm, resulting in a calculated electro-optic overlap integral $\Gamma$ of 0.06 while preserving maximum electric filed along the z-direction.

As shown in the fabrication process flow in Supplementary Figure 1, the metal electrodes are fabricated by two steps of lift-off processes on the $SiO_2$ cladding to ensure effective application of the electric field to the device via ground-signal-ground (GSG) probes. The change in waveguide effective refractive index due to the Pockels effect is manifested as a linear shift of the resonance peak. When the applied DC voltage is swept from 0 V to 30 V, the resonance wavelength is shifted by ~5.2 pm from the wavelength of 1544.1 nm (Fig. 3d). Therefore, the effective EO coefficient is extracted to be ~1.1 pm/V (see Methods)[27,31,32].

**High-frequency Characterization.** Utilizing the sideband measurement technique, we analyze the high-frequency response of the fabricated modulators. Sinusoidal RF signals of different frequencies are applied to the device, and the first-order modulation sidebands can be observed at corresponding frequencies separated from the carrier[33]. By measuring the power ratio of the main peak to the first sideband, the modulation index *m* can be extracted using Jacobi-Anger expansion of the modulated optical electrical field[34,35]. In the case where the $m \ll 1$, its zero-order and first-order Bessel functions can be approximated as $J_0(m) \sim 1$ and $J_1(m) \sim m/2$, respectively. Then, *m* can be extracted using the following equation:

$$P(mW) = \left[\frac{J_0(m)}{J_1(m)}\right]^2 = \left(\frac{2}{m}\right)^2 \tag{1}$$

When the operating wavelength is selected at a non-resonant point, the modulator can achieve effective intensity modulation. MRRs feature a Lorentzian transfer function $T(\theta)$, where $\theta$ denotes the round-trip phase shift. By selecting the operating wavelength at its maximum slope, the half-wave voltage $V_\pi$ can be effectively decreased by a factor of $1/2|dT/d\theta|_{max}$) compared to a single-arm-driven Mach–Zehnder (MZ) modulator with the same electro-optic interaction length $L$[36]. The optimal operating wavelength of the micro-ring modulator can thus be determined. We calculate the value of $2|dT/d\theta|$ by fitting the measured transmission profile, and subsequently extract the effective EO coefficient $r_{eff}$ and $V_\pi$ using Equation 2:

$$m = \frac{\pi V_{pp}}{V_\pi} = \frac{2\pi |dT/d\theta| V_{pp} n_o^4 r_{eff} \Gamma L}{\lambda g n_{eff}} \tag{2}$$

where $V_{pp}$ represents the amplitude of applied RF signal, and $n_{eff}$ is the effective refractive index of the waveguide. The experimental setup for sideband measurements is illustrated in Figure 4a (see Methods).

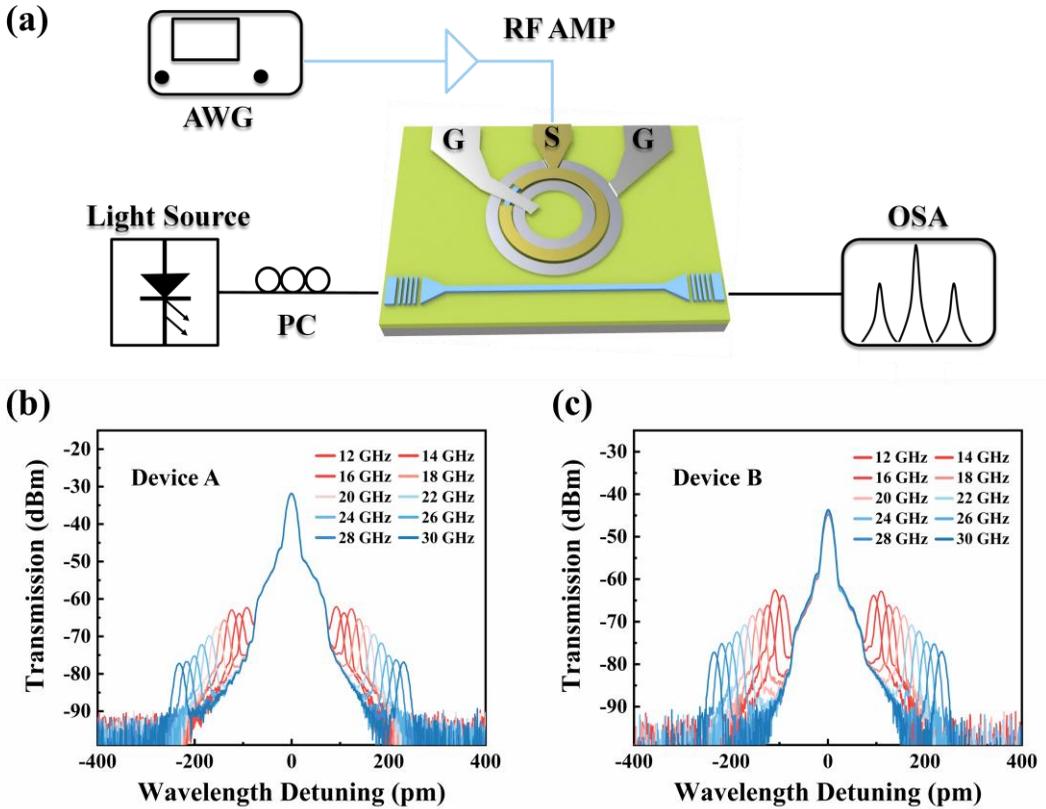

**Fig. 4 a** Schematic of the measurement setup characterizing the high-frequency EO response. Optical transmission spectra of ring modulators **b** A and **c** B measured at different RF frequencies. The measurements at 12 GHz are used to extract the effective in-device EO coefficient.

For evaluating the high-frequency EO performance of micro-ring modulators with various structures, two configurations are measured: an all-pass-type micro-ring (referred to as ring A) and an add-drop-type micro-ring (referred to as ring B), each having a radius of 40 μm. By applying RF signals ranging from 12 GHz to 30 GHz with the same $V_{pp}$ of 6.32 V to both devices, as shown in Figure 4b and Figure 4c, several pairs of sidebands are sequentially spaced with respect to the optical carrier, clearly demonstrating the generated intensity modulation. The effective in-device EO coefficient obtained for devices A and B is $r_{eff}$ = 2.47 pm/V and $r_{eff}$ = 2.86 pm/V, respectively. Furthermore, according to first-principles calculations, an increase in Sc concentration is predicted to yield a 2.5-times enhancement in the intrinsic EO coefficient of single-crystal $Al_{0.7}Sc_{0.3}N$ compared to $Al_{0.904}Sc_{0.096}N$ (Supplementary Figure 2), which is expected to further enhance the modulation efficiency.

Notably, EO coefficients measured at DC are slightly different from those measured at high frequencies. There are two factors may contribute to the observed phenomena. Firstly, the shielding effect: defects in the PECVD-$SiO_2$ cladding trap charges, altering the charge distribution of charges within the material and affecting the distribution of charges in the surrounding space. This alteration can lead to the shielding effect on charges located outside the oxide layer and impact the DC response of devices[37]. Secondly, there is a strong correlation between the EO coefficient and the frequency of the applied electric field. In the frequency range from DC to the first mechanical resonance (~10 MHz), the crystal is unclamped (stress-free). In this context, the combined influence of the electronic response, ionic response, and piezoelectric response jointly affects the value of the EO coefficient. At high frequencies ($10^6$ MHz > $f$ > $10^2$ MHz), the crystal is clamped (strain-free), and the influence of the piezoelectric response dissipates, leading to frequency-dependent variations in the EO coefficients[38,39]. It is noteworthy that some of the contributions are not necessarily independent and may work counter to one another.

Then, we analyze the RF modulation efficiency $V_\pi \cdot L$ of the micro-ring modulators. Four devices with different structures are selected and measured: device A (all-pass (AP), radius = 40 μm, Q = 15696, ER = 15 dB), B (add-drop (AD), radius = 40 μm, Q = 19580, ER = 27 dB), C (AP, radius = 100 μm, Q = 21410, ER = 14 dB), and D (AD, radius = 100 μm, Q = 18641, ER = 24 dB). As depicted in Figure 5a, device B shows a minimum

$V_\pi \cdot L$ of 3.12 V·cm at 14 GHz modulation frequency and the minimum $V_\pi \cdot L$ of devices A, C, and D is 8.81 V·cm, 17.37 V·cm and 14.6 V·cm, respectively.

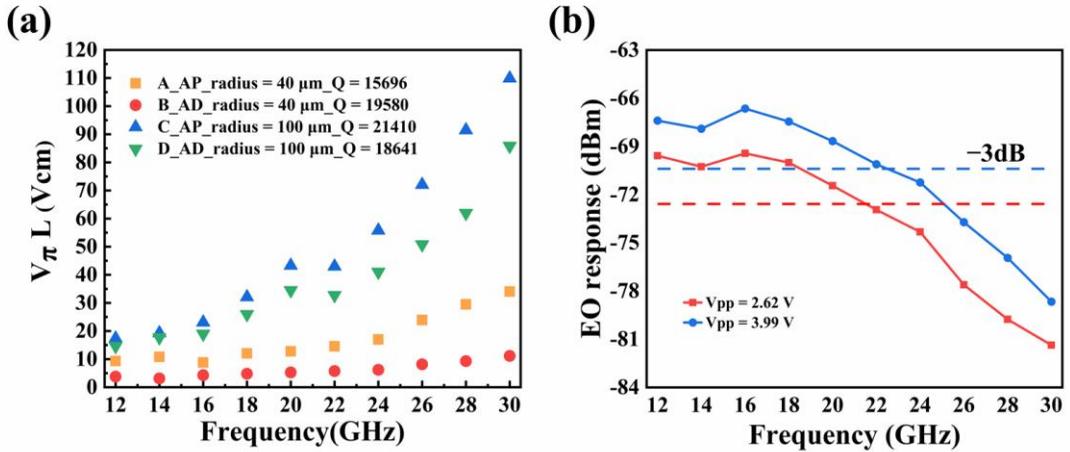

**Fig. 5 a** Measured half-wave voltage-length products at high frequencies. **b** The optical sideband power of device A versus various modulation frequencies and applied voltages.

The enhancement in modulation efficiency can be explained through straightforward metrics. Micro-ring modulators with larger ER (i.e., larger $|dT/d\theta|$) tend to possess higher modulation efficiency. This is due to the fact that a modulator with larger $|dT/d\theta|$ can generate a small resonant offset sufficient to achieve the desired modulation depth with a relatively low driving voltage. Also, high Q values promote adequate electro-optic interactions in modulators, thus yielding high modulation efficiencies. It is worth noting that, devices C and D have longer modulation lengths than devices A and B, leading to a more significant decline in their modulation efficiencies at higher frequencies. This can be attributed to the larger cavity, resulting in an increased capacitance that need to be driven by the applied source. To address this issue, potential improvements can be achieved through optimizing the design of traveling-wave electrodes and making slight increasement to electrode thickness. Additionally, modulation efficiency degrades with increasing RF frequencies across all four devices, primarily due to two factors: 1) Impedance decreases with increasing frequency ($\propto 1/|j\omega C|$), resulting in a reduced voltage drop across the EO modulation region; 2) Skin effect-induced ohmic loss becomes more severe at high frequency, further reducing the voltage drop across the modulation region[40].

The 3 dB EO bandwidth of our micro-ring modulators is determined by fitting the sideband powers at various applied RF frequencies starting from 12 GHz[36,41–43], primarily constrained by the resolving accuracy of the OSA. Figure 5b depicts the EO frequency response of device A under applied $V_{pp}$ of 3.99 V and 2.62 V, respectively. The 3 dB EO bandwidths are approximately 22 GHz. Generally, the EO bandwidth can be estimated using the simplified models: $1/f = 1/f_{RC} + 1/f_t$. The first term is relative to the RC time constant of the device and second term influenced by the lifetime of photons in the cavity. As the modulator is directly driven by the electric field applied to the electrodes with negligible resistance, $f_{RC}$ approaches infinity. Consequently, the cutoff frequency is mainly limited by the photon lifetime, expressed as $f_{3dB} = f_t \approx c/\lambda Q$, thereby restricting the maximum achievable Q-factor.

Here, optical peaking is experimentally observed, as shown in Figure 5b, which helps to extend the EO bandwidth beyond the photon lifetime limitation. The peaking phenomenon is related to transient responses in the optical domain and can be explained using a small-signal model based on perturbation theory[44–46]. The voltage applied to the micro-ring modulator induces changes in the refractive index, which further causes phase shifts. With the phase shift accumulates sufficiently, it switches from destructive to constructive interference between the resonant light circulating in the cavity and the input modulated light from the waveguide. This transient response leads to an overshoot in the output power, which can be observed as the peaking effect.

The modulation frequency at which the peaking occurs is related to the detuning between the resonant wavelength of the cavity $f_r$ and the optical carrier wavelength $f_0$. When the detuning $f_0 - f_r$ increases, the peaking occurs at higher modulation frequencies, resulting in a higher EO bandwidth (Supplementary S4). However, this enhancement comes at the cost of modulation efficiency, the EO response power decreases as the detuning increases, necessitating the use of amplifiers with higher gain in the test link. Therefore, we fixed a detuning of 0.195 nm, as this wavelength point yields the maximum $|dT/d\theta|$ for device A. As illustrated in Figure 5b, with the increase of $V_{pp}$, the overall power of the EO response shifts upwards. This phenomenon is attributed to a larger modulation voltage inducing a greater frequency shift in the resonator, which in turn amplifies the perturbation amplitude within the cavity. All curves were mathematically fitted, showing agreement with theoretical predictions (detailed fitting parameters and theoretical derivations are provided in Supplementary S4).

## Conclusions

In summary, we demonstrate the EO micro-ring modulator based on the AlScN-on insulator platform. The maximum effective in-device EO coefficient is 2.86 pm/V. At a modulation frequency of 14 GHz, the device also exhibits an RF modulation efficiency $V_\pi \cdot L$ of 3.12 V·cm. Moreover, the optical response of the modulator decays by 3 dB at a modulation frequency of 22 GHz. Leveraging the key advantages of AlScN thin films, which is the high-quality wafer-level deposition achieved through a low-cost fabrication process, fully CMOS-compatible AlScN modulators are poised to emerge as promising candidates for electro-optic signal processing on silicon-integrated photonics platforms.

## Methods

**Device fabrication.** Supplementary Figure 1 illustrate the whole fabrication process of the AlScN micro-ring modulator. To define the waveguide patterns, a 700-nm-thick silicon dioxide ($SiO_2$) mask layer is first deposited on top of the AlScN layer using plasma enhanced chemical vapor deposition (PECVD). After spin coating with Zep520 resist and lithography using electron beam lithography (EBL), the SiO$_2$ mask is etched by fluoride-based inductively coupled plasma (ICP) reactive ion etching (RIE). The remaining resist is then removed with acetone, and AlScN waveguide structures are etched with Cl$_2$/Ar chemistry-based ICP-RIE. The slight over-etching is performed to avoid increased light leakage caused by the unetched bottom corners of the waveguides. The etching selectivity of the resist to the hard mask and AlScN film to the hard mask is approximately 1:3.5 and 0.61:1, respectively. Finally, a 1.8-μm-thick silicon dioxide layer is deposited using PECVD.

Metal electrodes are fabricated on the SiO$_2$ cladding. The electrode fabrication process consists of two steps based on different critical dimensions (CD). Initially, electrode patterns are defined on the region of waveguide structures with smaller critical dimensions (CD) using EBL. Subsequently, a deposition process involving 15 nm of Ti and 300 nm of Au via electron beam evaporation is conducted, followed by a lift-off process to form the gold-colored regions. Then, laser direct writing is employed to define the metal pad patterns with larger CDs, followed by the deposition of 400 nm of Al and another lift-off process to form the gray-colored regions, as shown in Figure 1b in the main text. An overlap of approximately 6 μm between the two metal regions ensures the effective application of the electric field via ground−signal−ground (GSG) probe onto the device.

**EO coefficient extraction.** The ellipsoid equation of the AlScN crystal after applying the electric field $E$ can be simplified as follows:

$$\left(\frac{1}{n_o^2} + r_{13}E_z\right)x^2 + \left(\frac{1}{n_o^2} + r_{13}E_z\right)y^2 + \left(\frac{1}{n_e^2} + r_{33}E_z\right)z^2 + 2r_{42}E_y yz + 2r_{42}E_x xz = 1 \qquad (1)$$

where $E_x$ and $E_y$ represent the in-plane components of the electric field, $E_z$ denotes the out-of-plane component. The value of EO coefficients $r_{13}$ and $r_{33}$ in AlN are similar, while $r_{42}$ is observed to be quite small. To extract the $r_{13}$ and the $r_{33}$ of AlScN, the applied electric field must be parallel to the c-axis of the AlScN film, i.e., $E_z = E$, and $E_x = E_y = 0$. In this case, the change of the refractive index with respect to $E$ can be expressed as:

$$\Delta n_{x,y} = -\frac{1}{2} n_o^3 r_{13} E$$
$$\Delta n_z = -\frac{1}{2} n_e^3 r_{33} E \qquad (2)$$

$n_{x,y}$ denotes the effective refractive index of TE-polarized optical modes transmitted in the waveguide, while $n_z$ is the effective refractive index for TM modes. The voltage-induced effective refractive index change $\Delta n_{eff}$ is calculated from the measured resonant frequency drift by applying a DC voltage to the micro-ring modulator ($\Delta n_{eff}/n_g = \Delta\lambda/\lambda_0$), then the in-device effective EO coefficient of the AlScN waveguide at the operating wavelength is derived from $r_{eff} = 2n_{eff} g \Delta n_{eff} / n_o^4 V \Gamma$, where $V$ represents the applied DC voltage.

**Sideband measurement.** RF signals generated by an arbitrary waveform generator (AWG, Keysight M8194A) are amplified by an RF amplifier (AMP) before being applied to the device. The optical signal is directly coupled to the device from a tunable laser through an external polarization controller and a PM fiber. The modulated optical signal is then fed into an optical spectrum analyzer (OSA, Yokogawa AQ6370D), where the optical power of the sidebands observed at each RF frequency is recorded. To ensure the accuracy of measuring the EO response of the device, we calibrate the cables and maintain a relatively constant $V_{pp}$ at different frequencies.